\documentclass[prb,showpacs,byrevtex,altaffillsymbol,superscriptaddress,
citeautoscript,amsfonts,amssymb,amsmath,twocolumn,floatfix,altaffillsymbol]{revtex4}

\usepackage{graphicx}%
\usepackage{dcolumn}%
\usepackage{amsmath}%

\begin{document}

\date{\today}

\title{Crystal structure, electronic, and magnetic properties of the bilayered
rhodium oxide Sr$_3$Rh$_2$O$_7$}

\author{K. Yamaura}
\email[E-mail at:]{YAMAURA.Kazunari@nims.go.jp}
\homepage[Fax.:]{+81-298-58-5650}
\affiliation{Superconducting Materials Center, National Institute for Materials
Science, 1-1 Namiki, Tsukuba, Ibaraki 305-0044, Japan}

\author{Q. Huang}
\affiliation{NIST Center for Neutron Research, National Institute of Standards
and Technology, Gaithersburg, Maryland 20899}
\affiliation{Department of Materials and Nuclear Engineering, University of 
Maryland, College Park, Maryland 20742}

\author{D.P. Young}
\affiliation{Department of Physics and Astronomy, Louisiana State University,
Baton Rouge, LA 70803}

\author{Y. Noguchi}
\affiliation{Advanced Materials Laboratory, National Institute for Materials
Science, 1-1 Namiki, Tsukuba, Ibaraki 305-0044, Japan}

\author{E. Takayama-Muromachi}
\affiliation{Superconducting Materials Center, National Institute for Materials
Science, 1-1 Namiki, Tsukuba, Ibaraki 305-0044, Japan}

\begin{abstract}

The bilayered rhodium oxide Sr$_3$Rh$_2$O$_7$ was synthesized by high-pressure
and high-temperature heating techniques.
The single-phase polycrystalline sample of Sr$_3$Rh$_2$O$_7$ was 
characterized by measurements of magnetic susceptibility, electrical 
resistivity, specific heat, and thermopower.
The structural characteristics were investigated by powder neutron diffraction
study.
The rhodium oxide Sr$_3$Rh$_2$O$_7$ [{\it Bbcb}, {\it a} = 5.4744(8) \AA, {\it b} = 5.4716(9)
\AA, {\it c} = 20.875(2) \AA] is isostructural to the metamagnetic metal 
Sr$_3$Ru$_2$O$_7$, with five 4$d$ electrons per Rh, which is electronically 
equivalent to the hypothetic bilayered ruthenium oxide, where one electron per
Ru is doped into the Ru-327 unit.
The present data show the rhodium oxide Sr$_3$Rh$_2$O$_7$ to be metallic with enhanced 
paramagnetism, similar to Sr$_3$Ru$_2$O$_7$.
However, neither manifest contributions from spin fluctuations nor any traces of a 
metamagnetic transition were found within the studied range from 2 K to 390 
K below 70 kOe.

\end{abstract}


\maketitle

\section{Introduction}

The layered ruthenium oxides Sr$_2$RuO$_4$, Sr$_3$Ru$_2$O$_7$, and the 
perovskites SrRuO$_3$ and CaRuO$_3$, have attracted much attentions during the
last several years primarily because these systems allow quantum phase 
transitions in clean metals \cite{clean} to be addressed experimentally \cite 
{RMP01GRS,Cambridge99SS,RMP97SLS,PRB76JAH}.
The appearances of novel transitions associated with the correlated 4$d$ 
electrons continues to stir up significant issues within the condensed matter
community.
The layered member Sr$_2$RuO$_4$, for example, enters a superconducting state
at about 1 K in association with a rather unusual spin-triplet type 
electron-pairing, clearly at odds with predictions of conventional $s-$wave 
superconductivity \cite{PRB97KI,PRL01KI}.
The bilayered member, Sr$_3$Ru$_2$O$_7$, shows non-Fermi liquid transport in
the vicinity of a metamagnetic transition at high magnetic field and low 
temperature, suggesting that the scattering rate of the conduction electrons
may be influenced by quantum fluctuations, as opposed to more conventional
thermal fluctuations \cite{SCIENCE01SAG,PRL01RSP,PRB01DJS,PRB00SII,PRL98AVP}.
Finally, data suggest that a quantum critical point exists in the solid 
solution Sr$_{1-x}$Ca$_x$RuO$_3$ for $x\sim$0.7 between ferromagnetic 
(Sr-side) and nearly ferromagnetic (Ca-side) ordered states \cite 
{PRL99KY,JPSJ99TK,JPSJ98TK,PRB97GC,PRB01TH,PRB00TH}.

These materials have provided experimental opportunities to study criticality
of correlated electrons in the vicinity of the quantum critical points \cite
{SCIENCE01SAG,PRL01RSP,PRB01DJS,PRB00SII,PRL98AVP}. 
The present status of investigations of quantum phase transitions and 
criticality seem far behind that of the conventional transitions that are 
driven by thermal fluctuations \cite{RMP01GRS,Cambridge99SS,RMP97SLS,PRB76JAH}.
In order to promote progress in the studies of quantum phase transitions and
criticality in clean metals, and to answer many open questions aroused thus 
far, we need further opportunities to study these issues experimentally.
Searching for a new class of materials which show quantum critical behavior
within realizable magnetic, pressure, or chemical ranges, should be 
encouraged.
High-quality single crystals of these materials, if available, would be highly
desirable for experimental investigations.

We have been synthesizing the variety of compounds in the Sr-Rh(IV) oxide 
system, where Rh is expected to be 4$d^5 (t_{\text 2g}^5 e_{\text g}^0)$, by
applying high-temperature and high-pressure techniques at typically 6 GPa and
1500 $^\circ$C \cite{PRB01KY}.
Besides the counted members of the strontium-rhodium system \cite 
{JACS01KES,JAC00JRP,CM98JBC,JAC97RH}, the perovskite-type compound was 
recently obtained by quenching to room temperature at elevated pressure \cite
{PRB01KY}.
The formation of SrRhO$_3$ was an excellent example of what can be achieved by
high-pressure techniques -- no compounds at the ratio Sr:Rh = 1:1 had ever
been reported until the high-pressure experiment.
While the Sr substitution for LaRh$^{3+}$O$_3$ was studied by a regular 
solid-state-reaction technique at ambient pressure; it was then found that the
solubility limit of Sr at the La site was at most 10 \%, and the solid 
solution remained semiconducting in the very narrow region \cite{JSSC93TN}.
It is then clear that the formation of the 100 \% end-member SrRhO$_3$ goes far
beyond the limit achieved by the regular synthesis study, and it is indicative
of the dramatic effectiveness of the high-pressure experiment.
This synthesis technique can be expected to uncover further novel electronic
materials in addition to SrRhO$_3$.

The magnetic and electrical properties of SrRhO$_3$ were qualitatively similar
of those observed in the analogous ruthenium oxide CaRuO$_3$, the latter being
a nearly ordered ferromagnetic metal that is probably influenced by 
spin-fluctuations \cite{PRL99KY,JPSJ99TK,JPSJ98TK,PRB97GC,PRB01TH,PRB00TH}.
We had therefore expected unusual electronic properties in the unforeseen 
compounds in the rhodium oxide system, as well as the analogous ruthenium 
oxides.
We have focused our attention to obtaining further novel materials in the 
Sr-Rh-O space, and thus far, the Ruddlesden-Popper-type member 
Sr$_3$Rh$_2$O$_7$ was found.
In this paper we present data from neutron diffraction study, magnetic and 
electrical measurements on a polycrystalline sample of this new phase, and the
results are compared to the ruthenium analogue.

\section{Experimental}

\begin{figure}
\includegraphics[width=8cm]{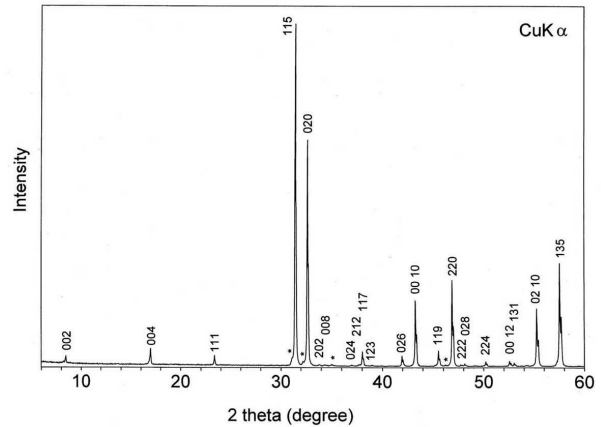}
\caption{Powder x-ray diffraction profile (CuK$\alpha$) of Sr$_3$Rh$_2$O$_7$,
obtained at room temperature. A tentatively applied quasi-tetragonal unit cell
with $a =$ 5.474(1) \AA~and $c =$ 20.88(1) \AA~yields a clear assignment of 
($hkl$) numbers to almost all peaks, indicating quality of the sample. 
Unindexed peaks are marked by stars.}
\label{fig1}
\end{figure}

The polycrystalline sample of Sr$_3$Rh$_2$O$_7$ was prepared as follows. The
fine and pure ($\ge99.9\%$) powders of SrO$_2$, Rh$_2$O$_3$, and Rh were mixed
at the compositions Sr$_3$Rh$_2$O$_{7+z}$, where $z =$ 0.0 and 0.1.
Approximately 0.2 grams of each stoichiometry was placed into a platinum 
capsule and then compressed at 6 GPa in a high-pressure apparatus, which was
originally developed in our laboratory \cite{press}.
The sample was then heated at 1500 $^\circ$C for 1 hr and quenched to room 
temperature before releasing the pressure.
The quality of the products was investigated by means of powder x-ray 
diffraction (CuK$\alpha$).
The x-ray profile was carefully obtained at room temperature using the 
high-resolution-powder diffractometer (RINT-2000 system, developed by RIGAKU,
CO), which was equipped with a graphite monochromator on the counter side.
We found that the both samples were of high quality; there were no significant
impurities in either product, and no remarkable difference between them. The
x-ray pattern for the powder sample at $z =$ 0.0 is shown in Fig.\ref{fig1}.
A quasi-tetragonal cell with $a =$ 5.474(1) {\AA} and $c =$ 20.88(1) {\AA} was
tentatively applied to the data, and a clear assignment of ($hkl$) numbers 
to all the major peaks was obtained, indicating the high quality of the 
sample. 
The one probable impurity phase (less than 1\%) was SrRhO$_3$ \cite{PRB01KY}.
The major phase in the samples could be reasonably assigned to 
Sr$_3$Rh$_2$O$_7$, of which there are no known records in the literature.

Possible oxygen vacancies in the compound were quantitatively investigated in
detail by thermogravimetric analysis.
Approximately 30 milligrams of the $z =$ 0.0 powder was slowly heated at 5 
$^\circ$C/min in a gas mixture (3\% hydrogen in argon) and held at 800 
$^\circ$C until the weight reduction became saturated.
Calculated oxygen composition from the weight loss data was 7.01 per formula
unit, indicating good oxygen stoichiometry.

Another sample of Sr$_3$Rh$_2$O$_7$, prepared at identical synthesis 
condition, was used for structural study by means of neutron diffraction.
The powder neutron data of powder sample of Sr$_3$Rh$_2$O$_7$ ($\sim$0.2 
grams) were collected at room temperature using the BT-1 high-resolution 
powder diffractometer at the NIST Center for Neutron Research, employing a 
Cu(311) monochromator to produce a monochromatic neutron beam of wavelength 
1.5396 \AA.
Collimators with horizontal divergences of 15$^\prime$, 20$^\prime$, and 
7$^\prime$ of arc were used before and after the monochromator, and after the
sample, respectively.
The intensities were measured in steps of 0.05$^\circ$ in the 2$\theta$ range
3$^\circ$--168$^\circ$. 
The structural parameters were refined using the program GSAS \cite{GSAS}.   
The neutron scattering amplitudes used in the refinements were 0.702, 0.593,
and 0.581 ($\times$10$^{-12}$ cm) for Sr, Rh, and O, respectively.

The magnetic properties of the sample ($z=$0.0) were studied in a commercial
apparatus (Quantum Design, MPMS-XL system) between 2 K and 390 K.
The magnetic susceptibility data were collected at 1 kOe and 70 kOe, and 
magnetization curves were recorded between -70 kOe and 70 kOe after cooling 
the sample at each 2 K, 5 K and 50 K.
The electrical resistivity was measured by a conventional ac-four-terminal 
technique at zero field and at 70 kOe.
The ac-gauge current was 0.1 mA at 30 Hz. 
The selected sample ($z=$0.0) was cut out into a bar shape, and the each face
was polished with aluminum-oxide-lapping film.
In order to decrease the contact resistance with the sample, gold pads ($\sim$
200 nm in thickness) were deposited at four locations along the bar, and 
silver epoxy was used to fix platinum wires ($\sim$ 30 $\mu$m$\phi$) at each
gold terminal.
The contact resistance at the four terminals was less than 8 ohm. 

\section{Results and Discussions}

\begin{table*}
\caption{Structure parameters of Sr$_3$Rh$_2$O$_7$ at room temperature. Space
group: {\it Bbcb} (No.68). The lattice parameters are {\it a} = 5.4744(8) \AA,
{\it b} = 5.4716(9) \AA, and {\it c} = 20.875(2) \AA.
The volume of the orthorhombic unit cell is 625.3(2) \AA$^3$. $Z$= 4.
The calculated density is 6.17 g/cm$^3$.
The constraints on the analysis: $y$[O(3)]$=-x$[O(3)]+1/2, $x$[O(3')]
$=1-x$[O(3)], $y$[O(3')]$=-y$[O(3)], $z$[O(3')]$=z$[O(3)], $B$[O(3')]$=B$[O(3)], and
$n$[O(3')]$=1-n$[O(3)]. 
The $R$ factors were 5.42\% ($R_{\rm p}$) and 6.47\% ($R_{\rm wp}$). 
Selected bond distances (\AA) and angles ($^\circ$) are shown in the bottom part. 
}
\label{table1}
\begin{ruledtabular}
\begin{tabular}{lllllll}
Atom &Site &$x$         &$y$          &$z$         &$B$(\AA$^2$)&$n$ \\
\hline
Sr(1)&$4a$  &1/4         &1/4         &0           &0.5(2)    &1     \\
Sr(2)&$8e$  &1/4         &1/4         &0.1867(3)   &0.8(2)    &1     \\
Rh   &$8e$  &1/4         &1/4         &0.4013(4)   &0.1(2)    &1     \\
O(1) &$4b$  &1/4         &1/4         &1/2         &1.2(3)    &1     \\
O(2) &$8e$  &1/4         &1/4         &0.3062(4)   &0.6(2)    &1     \\
O(3) &$16i$ &0.5473(5)   &-0.0473(5)  &0.0973(4)   &0.6(1)    &0.92(2)\\
O(3')&$16i$ &0.4527(5)   &0.0473(5)   &0.0973(4)   &0.6(1)    &0.08(2)\\
\hline
Sr(1)--O(1)   &$\times$2   &2.7358(4) &&Rh--O(1)      &  	 &2.060(8)  \\
Sr(1)--O(1)   &$\times$2   &2.7372(4) &&Rh--O(2)      &	         &1.985(12) \\
Sr(1)--O(3)   &$\times$4   &3.069(7)  &&Rh--O(3)      &$\times$2 &1.9694(8) \\
Sr(1)--O(3)   &$\times$4   &2.566(8)  &&Rh--O(3)      &$\times$2 &1.9697(8) \\
Sr(2)--O(2)   &            &2.494(11) &&O(1)--Rh--O(3)&          &89.1(4)   \\
Sr(2)--O(2)   &$\times$2   &2.7411(7) &&O(2)--Rh--O(3)&          &90.9(4)   \\
Sr(2)--O(2)   &$\times$2   &2.7398(7) &&O(1)--Rh--O(2)&          &180       \\
Sr(2)--O(3)   &$\times$2   &2.44(1)   &&Rh--O(3)--Rh  &          &158.5(2)  \\
Sr(2)--O(3)   &$\times$2   &2.964(8)  &&              &          &          \\
\end{tabular}
\end{ruledtabular}
\end{table*}

The crystal structure of the title compound was investigated by means of 
neutron powder diffraction at room temperature.
In an effort to maximize the homogeneity of the sample, only one pellet was 
subjected to the neutron study rather than a mixture of multiple pellets, even
though the total sample mass was far below the regular level.
The degree of local structure distortions, such as cooperative rotations of 
metal-oxygen octahedra and frequency of stacking faults in the layered 
structure, may depend sensitively on the synthesis conditions.
Technical problems, such as small sample volume within the high-pressure 
apparatus and uncertainties in the maximum temperature during the 
heating process, could contribute to small variations in sample quality.
Measuring an individual pellet by neutron powder diffraction was intended to
preclude possible errors in the analysis of the local structural distortion.
The structural distortion was naturally expected in Sr$_3$Rh$_2$O$_7$ based 
on the analogy with Sr$_3$Ru$_2$O$_7$, since the ionic size of $^{\rm VI}$Rh$^
{4+}$ (0.60 \AA) \cite{ACA76RDA} is very close to that of $^{\rm VI}$Ru$^{4+}
$ (0.62 \AA) \cite {JSSC00HS,PRB00HS,PRB98QH}.
Furthermore, the expected distortions were hard to detect with normal x-ray 
diffraction.

\begin{figure}
\includegraphics[width=8cm]{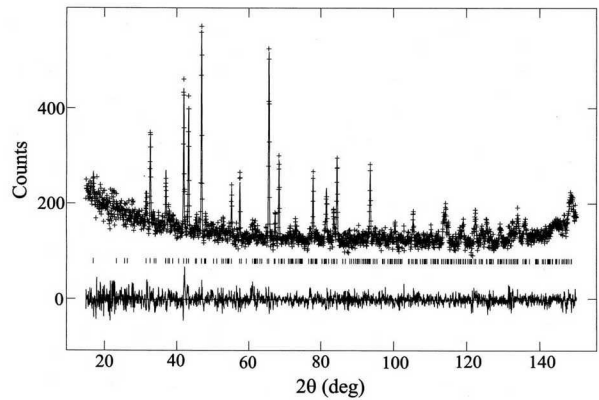}
\caption{Neutron diffraction profile of the powder sample ($\sim$0.2 grams) 
of Sr$_3$Rh$_2$O$_7$, obtained at room temperature at $\lambda=$ 1.5396 \AA.
Vertical bars indicate expected peak positions for the orthorhombic ({\it 
Bbcb}) structure model. The difference between the orthorhombic model (solid
lines) and the data (crosses) is shown below the bars column. Although 
intensity statistic appeared to be rather lower due to the small sample mass,
the analysis succeeded in obtaining a high-quality solution.}
\label{fig2}
\end{figure}

The pellet as made in a platinum capsule was highly polished with a fine 
sandpaper to reduce possible Pt contamination. 
The pellet ($\sim$0.2 grams) was then finely ground for the neutron 
diffraction study.
The neutron data were collected at room temperature for approximately one and
a half days; the profile is shown in Fig.\ref{fig2}.
Although the sample powder was exposed to the neutron beam for many hours, the
intensity did not reach normal levels due to the small sample mass.
It was, however, sufficient enough to deduce structural information from the 
data.
The small impurity contribution, found in the x-ray study, and possible 
magnetic contributions were not investigated due to the low intensity level.

\begin{figure}
\includegraphics[width=8cm]{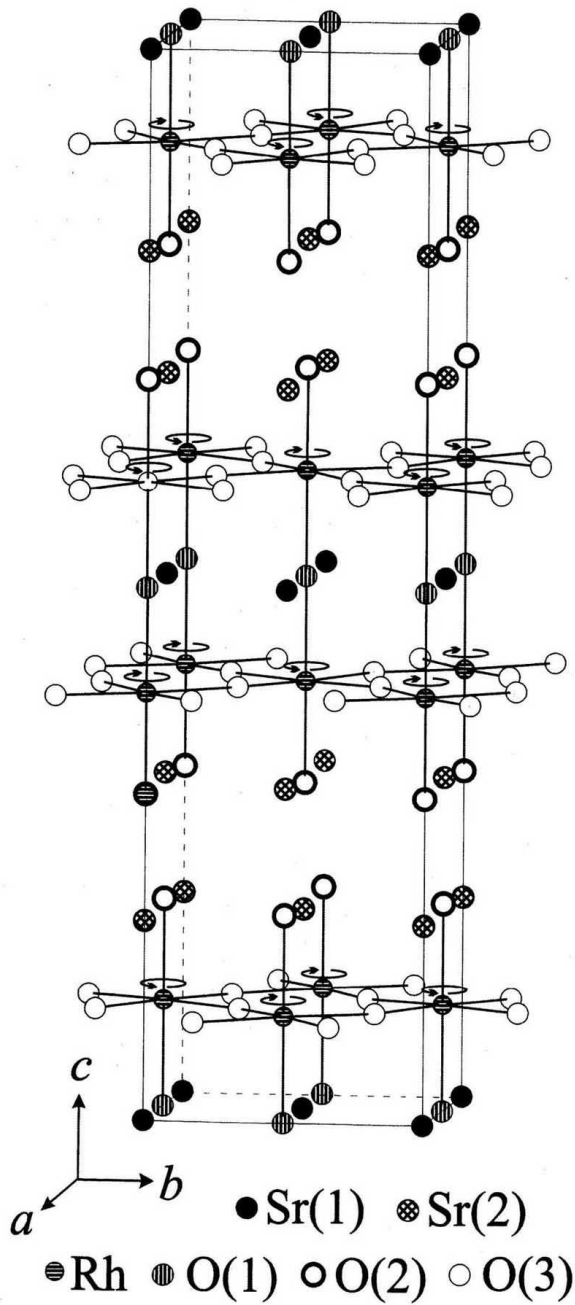}
\caption{Structural view of Sr$_3$Rh$_2$O$_7$. Fine lines signifies the 
orthorhombic unit cell. Cooperative rotations along $c-$axis of the RhO$_6$ 
octahedra are indicated by circular arrows.}
\label{fig3}
\end{figure}

The major concern in the structure investigation of the title compound was 
whether a cooperative rotation of the metal-oxygen octahedra occurs as is 
found in the analogous Sr$_3$Ru$_2$O$_7$ \cite{JSSC00HS,PRB00HS,PRB98QH}.
In the study of the $ruthenium$ oxide by means of neutron diffraction, eight
independent modes for the cooperative rotation 
($Bbcm$, $Bbcb$, $Bbmm$, $Bmcm$, $P4_2/mnm$, $P4_2/mcm$, $P4_2/m$, $B112/m$), 
combinations of some of those, and the rotation-free mode ($I4/mmm$) were 
tested.
It was found that the $Bbcb$ mode resulted in the highest quality of the 
profile analysis \cite{JSSC00HS}.
Here, in the analysis of the profile of Sr$_3$Rh$_2$O$_7$, we first applied 
the $Bbcb$ mode and found it to produce reasonable solutions ($R_{\rm p}=$ 
5.42\%, $R_{\rm wp} =$ 6.47\%) to describe the cooperative rotation of RhO$_6$
octahedra in Sr$_3$Rh$_2$O$_7$.
Although the $Bbcb$ mode is the most likely structural distortion in 
Sr$_3$Ru$_2$O$_7$, other lower symmetry modes, as mentioned above, are still
possible.
Unfortunately, the low intensity of the data in the present study made it 
difficult to completely dismiss the other modes.
Detailed analysis of the issue is left for future work, in which further 
progress of the high-pressure-synthesis technique will allow us to obtain a 
larger amount of highly homogenized sample.

Probable stacking faults were examined by introducing the O(3') elements with
some constraints (caption of Table \ref{table1}), as was done for 
Sr$_3$Ru$_2$O$_7$. 
As a result, significant improvements in the overall fit were obtained 
($<\sim$2\% in $R$ values).
The estimated occupancy of O(3') was $\sim$8\%, indicating the frequency of 
the stacking faults.
The estimated structural parameters of Sr$_3$Rh$_2$O$_7$, and selected bond 
distance and angles, are arranged in Table \ref{table1}.
The isotropic atomic displacement parameter ($B$) of the rhodium ion 
(8$e$--site) is rather unusual, possibly because the low level of the 
intensity statistics reduced reliability of the thermal parameters somewhat.
We believe further structure studies, on not only a larger amount of sample 
powder, but also a single crystal of Sr$_3$Rh$_2$O$_7$, might provide much 
more reliable data.

The structure of Sr$_3$Rh$_2$O$_7$ is shown in Fig.\ref{fig3}. 
The directions of rotation are indicated by circular arrows. 
The rotation results in the bond angle of Rh--O(3)--Rh being 159 degrees 
rather than 180 degrees for an ideally rotation-free structure.
The rotation angle of RhO$_6$ octahedra along $c-$axis is 10.5 degrees, which
is slightly greater than that of RuO$_6$ octahedra in the analogous 
Sr$_3$Ru$_2$O$_7$ (6.8 degrees) \cite{JSSC00HS}.
A general diagram for the degree of distortion in Ruddlesden-Popper-type 
compounds $A_3M_2$O$_7$ ($A:$ alkaline and/or rare earth metal; $M:$ 
transition metal) was proposed, in which the real members of $A_3M_2$O$_7$ 
were divided into two categories: one with and one without distortions 
associated with cooperative rotations of the $M$O$_6$ octahedra \cite
{JSSC00HS}.
A criterion named ``zero strain line'' was established for judging which
category each member should belong to.
The combination of radii of $^{\rm VI}$Rh$^{4+}$ (0.60 \AA) and $^{\rm XII} 
$Sr$^{2+}$ (1.44 \AA) in Sr$_3$Rh$_2$O$_7$ \cite{ACA76RDA} yields a point 
below the ``zero strain line'' in the proposed diagram, implying that 
Sr$_3$Rh$_2$O$_7$ should be in the distortion-free category.
This classification for the Sr$_3$Rh$_2$O$_7$ case appears to be incorrect 
as dictated by the data in the present study \cite{JSSC00HS}.
Characteristics of chemical bonds may play a pivotal role in the distortion 
process, and their consideration may be necessary if a complete diagram is to
be developed.

\begin{figure}
\includegraphics[width=8cm]{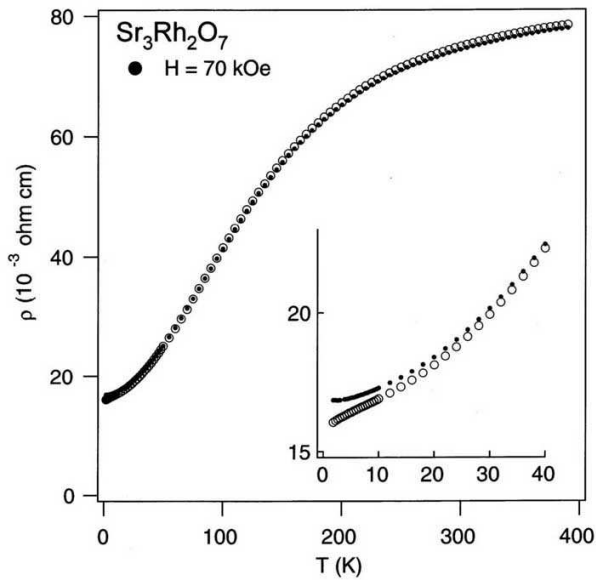}
\caption{Temperature dependence of the electrical resistivity of the 
polycrystalline Sr$_3$Rh$_2$O$_7$ at zero filed and 70 kOe. 
Inset shows an expansion of the low-temperature portion.}
\label{fig4}
\end{figure}

The electrical resistivity data of the polycrystalline sample of 
Sr$_3$Rh$_2$O$_7$ are shown in Fig.\ref{fig4}.
Temperature and magnetic field dependence was studied between 2 K and 390 K. 
The resistivity at room temperature is $\sim$75 m$\Omega$-cm, which is 
approximately one magnitude lager than that of the 3D SrRhO$_3$ \cite{PRB01KY}
and one magnitude smaller than that of the 2D Sr$_2$RhO$_4$, which is a barely
metallic conductor \cite{JSSC95MI,PRB94TS}. 
The series of the rhodium oxides Sr$_{n+1}$Rh$_n$O$_{3n+1}$ ($n=$ 1, 2, and 
$\infty$) consist of solely homovalent elements including Rh$^{4+}$ (4$d^5:$
may be in the low-spin configuration $t_{\rm 2g}^5 e_{\rm g}^0$), and all of
them clearly show metallic character, while the electronic dimensionality 
varies from two to three.
This suggests that the metallic character is developed with increasing 
electronic dimensionality.
Further studies involving band structure calculations could be significant in
revealing how essential the dimensionality is to the electronic transport of
the rhodium oxide series.
A small magnetic field dependence was found at low temperature ($<\sim20$K)
as shown in the inset of Fig.\ref{fig4}.
Future studies on single crystals may provide insight into the mechanism 
responsible for the positive magnetoresistance.

\begin{figure}
\includegraphics[width=8cm]{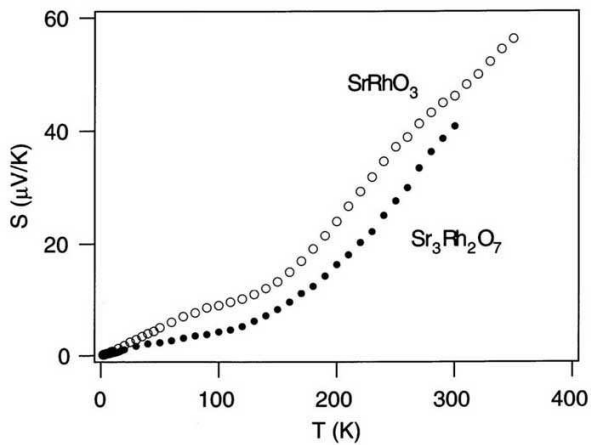}
\caption{Thermoelectric property of the polycrystalline Sr$_3$Rh$_2$O$_7$ and
SrRhO$_3$. The data for the perovskite compound were taken from a previous 
report \cite{MRS02KY}.}
\label{fig5}
\end{figure}

The Seebeck coefficient of the title compound was measured and compared with
that of the perovskite SrRhO$_3$ (Fig.\ref{fig5}) \cite{MRS02KY}.
The Seebeck coefficient of Sr$_2$RhO$_4$, reported previously, was 
approximately $+40 \mu$V/K at room temperature and decreases linearly to 
approach zero upon cooling \cite{PRB94TS}.
The data of all three compounds (Sr$_2$RhO$_4$, Sr$_3$Rh$_2$O$_7$, and 
SrRhO$_3$) are qualitatively consistent with the metallic conductivity 
observed for each, and indicate that the sign of the majority carriers is 
positive in the metallic conducting state.
Among the Seebeck coefficient of the three compounds, it is hard to find a 
remarkable difference qualitatively and quantitatively, in contrast to the 
case of the resistivity data.
Although a small change in slop in the Seebeck coefficient was observed in 
both Sr$_3$Rh$_2$O$_7$ and SrRhO$_3$ at approximately 150 K (Fig.\ref{fig5}), 
there were, however, no corresponding anomalies in other transport data 
around that temperature.
The change may be structurally related, and further studies of the temperature
dependence of structural distortions may help to address the issue.

\begin{figure}
\includegraphics[width=8cm]{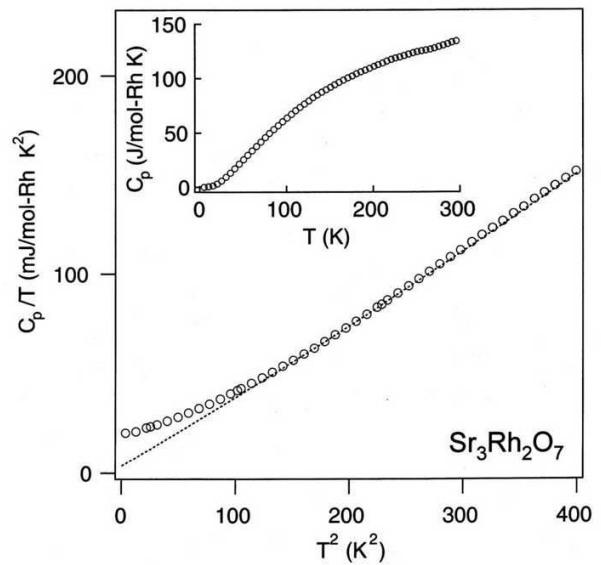}
\caption{Specific heat of the polycrystalline Sr$_3$Rh$_2$O$_7$. The data are 
plotted in $C_{\rm p}/T$ vs $T^2$ (main panel) and $C_{\rm p}$ vs $T$ (inset)
forms. The dotted curve represents a fit to the data (details are in the text).}
\label{fig6}
\end{figure}

The specific heat data are plotted as $C_{\rm p}/T$ vs $T^2$ (main panel) and
$C_{\rm p}$ vs $T$ (inset) in Fig.\ref{fig6}.
The electronic and lattice contributions to the specific heat can be isolated
from the data in the main panel by using the following low-temperature form 
($T \ll \Theta_{\rm D}$):
\begin{eqnarray*}
{C_{\text v} \over T} = \gamma + {12\pi^4 \over 5}Nk_{\text B}\Big({1 \over 
\Theta_{\text D}^3}\Big)T^2,
\label{}
\end{eqnarray*}
where $\Theta_{\rm D}$, $\gamma$, $\beta$, $k_{\rm B}$, and $N$ are Debye 
temperature, Sommerfeld's coefficient, Boltzmann's constant, and Avogadro's 
number, respectively.
The two independent parameters, $\Theta_{\rm D}$ and $\gamma$, which indicate
physical characteristics of the compound, were estimated by a least-square 
method in the linear range between 150 K$^2$ and 450 K$^2$; we obtained 
$\Theta_{\rm D} \sim$ 172 K and $\gamma \sim$ -0.00199 
J$\cdot$mol-Rh$^{-1}\cdot$K$^{-2}$.
The estimation of both $\gamma$ and $\Theta_{\rm D}$ was, however, not solid; 
a small degree of variations was found.
For example, when a $T^4$ term was added to the equation, the fits achieved 
were $\Theta_{\rm D} $ of $\sim$ 180 K, $\gamma$ of $\sim$0.00365 
J$\cdot$mol-Rh$^{-1}\cdot$K$^{-2} $, and $\beta$ (the coefficient of the $T^4$
term) of 7.93$\times$10$^{-8}$ J$\cdot$mol-Rh$^{-1}\cdot$K$^{-6}$, as indicated
by the dotted curve.
Although it is technically difficult to determine a precise value for 
$\gamma$, it should, however, be very small in contrast to that of the 
metallic SrRhO$_3$ ($\gamma$ = 7.6 mJ/mol-Rh/K$^2$) \cite{PRB01KY}.
The depressed metallic character of the bilayered Sr$_3$Rh$_2$O$_7$ is 
reflective in the very small value of $\gamma$.
It would be interesting to compare the value of $\gamma_{\rm band}$, computed
from the density of states at the Fermi level by means of band-structure 
calculations, with and without consideration of spin polarization.
In the above analysis we assumed that $C_{\text p}$ was approximately equal to
$C_{\text v}$ because it was in the low temperature limit. 

As expected from the Debye temperature (170 K--180 K) of the layered oxide 
Sr$_3$Rh$_2$O$_7$, even within the studied temperature range, the specific 
heat approaches the high-temperature limit, roughly calculated to be $\sim$150
J$\cdot$mol-Rh$^{-1}\cdot$K$^{-1}$ [6(atoms per unit cell) $\times$3 
(dimensionality per atom) $\times k_{\rm B}$N(Boltzmann and Avogadro's 
constants)].
The Debye temperature of Sr$_3$Rh$_2$O$_7$ is close to that of the perovskite
SrRhO$_3$ ($\Theta_{\rm D} \sim$ 190 K).
To our knowledge, the Debye temperature of Sr$_2$RhO$_4$ has not been 
reported.

A small component, which appears in the low-temperature limit of the $C_{\text
p}/T$ vs $T^2$ curve (the data points gradually apart from the fitting curve
on cooling), is probably due to a magnetic origin.
In order to make clear the origin of the probable magnetic component, several
magnetic models were tested on the specific heat data.
The studies, however, did not yield a clear assignment of magnetic model to 
the origin because there were still too many possibilities related to a nuclear 
Schottky specific heat, an empirical temperature-independent term, ferro- and
antiferromagnetic spin fluctuations, and associations of some of those.
As far as the present investigation goes, it was therefore unlikely to 
establish a clear picture with a reasonable sense to understand the appearance
of the very small component in the low-temperature specific heat data.

\begin{figure}
\includegraphics[width=8cm]{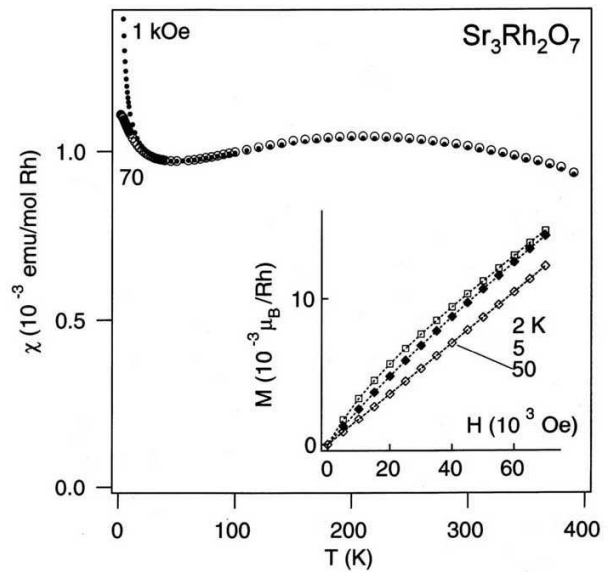}
\caption{Temperature dependence of the magnetic susceptibility of the 
polycrystalline Sr$_3$Rh$_2$O$_7$ at 1 kOe and 70 kOe on cooling, and applied
magnetic field dependence of the magnetization (inset) at 2 K, 5 K and 50 K.}
\label{fig7}
\end{figure}

Nearly-temperature-independent magnetic susceptibility data, measured between
2 K and 390 K, are shown in Fig.\ref{fig7}.
The majority of the data could be characterized as Pauli paramagnetic with an
enhanced $\chi(0)$.
The $\chi(300)$ is approximately 1.0$\times10^{-3}$ emu/mol-Rh, 
nearly one magnitude larger than those of normal paramagnetic metals,
including Rh(IV)O$_2$ \cite{PPS61HK,PPS60DWB,PPS51FEH}.
This is a curious result, considering the very small $\gamma$-value 
estimated from the specific heat measurements. 
More will be said on this below. 
Alternatively, a layered spin system, despite the metallic character, is 
probable to account for the magnetic characteristics with the broad maximum at 
about 200 K as is the case for the Sr doped La$_2$CuO$_4$ \cite{RMP98MI}.
Further studies to explore probable localized magnetic moments should be of 
interest.
The steep upturn in low temperature at 1 kOe is due to a small concentration
of magnetic impurities, which could be well fit to a Curie-Weiss law; the 
magnetic parameters were $C=$ 0.00194 emu K/mol-Rh (Curie constant), $\theta_
{\rm W} =$ -0.1462 K (Weiss temperature), and $\chi_0 =$ 9.21$\times10^{-4}$
emu/mol-Rh (temperature-independent susceptibility).
The rather small value of $\theta_{\rm W}$, less than 1 K, indicates that the
diluted moments are free from magnetic interactions, as is often the cases.
The $M$ vs $H$ curves measured at 2 K, 5 K and 50 K are shown in the inset of
Fig.\ref {fig7}.
The small component ($\sim$0.001$\mu_{\rm B}$ at 70 kOe and 5 K), which is 
superimposed on the paramagnetic background, is probably due to the magnetic
impurity outlined above.
While a metamagnetic transition was clearly found in the analogous 
Sr$_3$Ru$_2$O$_7$ at $\sim$ 50 kOe and 2.8 K in fields both parallel and 
perpendicular to the $c-$axis of a single crystal \cite{PRL01RSP}, no trace 
of one could be seen in the data of Sr$_3$Rh$_2$O$_7$ even at 2.0 K.
Such a transition may be shifted to higher fields in the Rh sample, and in 
order to investigate this possibility, a pulsed high-magnetic-field study at
low temperature is scheduled for Sr$_3$Rh$_2$O$_7$.

\section{Conclusions}

High-pressure synthesis techniques have been utilized in an effort to add to
the variety of experimental systems currently being used in the study of 
quantum phase transitions and criticality.
The systematic synthesis experiments performed thus far on the Rh$^{4+}$ 
(4$d^5:t_{\rm 2g}^5 e_{\rm g}^0$, expected) oxide system have revealed a 
new class of Ruddlesden-Popper phases Sr$_{n+1}$Rh$_{n}$O$_{3n+1}$ ($n=$ 1 
[ref. 31 and 32], 2 [this work], and $\infty$ [ref. 19]).
The neutron diffraction study clearly reveals that the new metallic rhodium 
oxide, Sr$_3$Rh$_2$O$_7$, is isostructural to the ruthenium oxide 
Sr$_3$Ru$_2$O$_7$, which has received much recent attention due to 
its intriguing quantum characteristics. 
The magnetic and transport data presented here, however, do not indicate 
clear contributions from quantum fluctuations in the studied range of 
temperature and magnetic field, in contrast to what was observed in the 
ruthenium compound \cite{PRB98SI,JSSC95RJC}.
Since the Curie-Weiss-like behavior was not seen (except for the impurity term)
even in the high-temperature portion of the magnetic susceptibility data, the
self-consistent renormalization theory of spin fluctuations for both 
antiferro- and ferromagnetic nearly-ordered magnetic metals may not be 
applicable to the present magnetic and specific heat data \cite 
{Springer85TM,AP00TM}.

While the magnetic susceptibility data of Sr$_3$Rh$_2$O$_7$ suggest a rather
large density of states at the Fermi level (similar to what was observed in 
SrRhO$_3$), the linear term in the heat capacity is, however, remarkably small -- on the
order of semimetals like Bi and Sb.
The resistivity is about 10 times larger than that of SrRhO$_3$, qualitatively
consistent with the rather small $\gamma$.
It is possible that the influence of disorder in the polycrystalline sample at a
level undetectable by the techniques in the present study masks the intrinsic
nature of ideally clean Sr$_3$Rh$_2$O$_7$.
In the studies of the ruthenium oxides, for example, the quantum properties 
were found to be rather fragile to such disorder \cite{PRL02LC,PRL98APM}.
Further information, including carrier density, degree of disorder, band
structure, would prove useful in clearly identifying the intrinsic properties 
of Sr$_3$Rh$_2$O$_7$.
High-quality single crystals of sufficient size for electronic transport 
studies synthesized in the high-pressure cell would afford the opportunity of 
further investigations of Sr$_3$Rh$_2$O$_7$ at a much deeper level.

In the ruthenium oxide family, much effort have been spent in searching for 
new superconducting compounds, but only the 214 phase is a superconductor. 
One member found recently (non-superconducting) is the double-layered 
Sr$_3$Ru$_2$O$_7$F$_2$, where Ru ions are pentavalent \cite{PRB00RKL}.
One electronic hole per formula unit was formally doped into metallic 
Sr$_3$Ru$_2$O$_7$, and an ordered magnetic moment ($\sim$1$\mu_{\text B}$ per
Ru) was found below about 185 K.
On the other hand, Sr$_3$Rh$_2$O$_7$ has five 4$d$ electrons per formula 
unit, i.e. one electron per Ru is hypothetically doped into Sr$_3$Ru$_2$O$_7$,
and no localized moments observed.
The following is a general description of the three bilayered compounds: The
localized character at the 4$d^3$ layered compound gets weakened at 4$d^4$ and
becomes much more itinerant at 4$d^5$.
The overall comprehensive picture about the electronic structure of various 
4$d$ compounds has probably not yet been established.
Further studies, including theoretical considerations and a systematic study
of isovalent and aliovalent chemical doping to the series Sr$_{n+1}$Rh$_n$O$_
{3n+1}$ would be of interest.
Further progress of the high-pressure-synthesis technique may allow future 
investigations of new and unforeseen electronic materials those are potentially
intriguing.

\acknowledgments
We wish to thank Dr. M. Akaishi (NIMS) and Dr. S. Yamaoka (NIMS) for their 
advice on the high-pressure experiments. This research was supported in part
by Superconducting Materials Research Project, administrated by the Ministry
of Education, Culture, Sports, Science and Technology of Japan. One of us 
(K.Y.) was supported by the Domestic Research Fellowship, administrated by the
Japan Society for the Promotion of Science.


\pagebreak

\end{document}